\documentstyle[twoside]{article}
\newcommand{\la}[1]{\mb{\lambda}^{\rm #1}}
\newcommand{\cl}{\hspace{-1ex}&\circa{<}\hspace{-1ex}&}
\renewcommand{\Im}{\mathop{\rm Im}}

\newcommand{\GeV}{\,{\rm GeV}}
\newcommand{\TeV}{\,{\rm TeV}}
\newcommand{\eV}{\,{\rm eV}}
\newcommand{\NP}{Nucl. Phys.}
\newcommand{\PRL}{Phys. Rev. Lett.}
\newcommand{\PL}{Phys. Lett.}
\newcommand{\PR}{Phys. Rev.}

\newcommand{\eq}[1]{~{\rm (\ref{eq:#1})}}
\newcommand{\md}[1]{\langle#1\rangle}
\newcommand{\MGUT}{M_{\rm G}}
\def\Red{}
\def\Black{}
\def\Blue{}

\newcommand{\lascia}[1]{}
\makeatletter
\def\art{\@ifnextchar[{\eart}{\oart}}
\def\eart[#1]#2#3#4#5#6{{\rm #2}, {\em #3 \bf #4} {\rm (#6) #5 [#1]}}
\def\hepart[#1]#2{{\rm #2, \em#1}}
\newcommand{\oart}[5]{{\rm #1}, {\em #2 \bf #3} {\rm (#5) #4}}
\newcommand{\y}{{\rm and} }

\newcounter{alphaequation}[equation]
\def\thealphaequation{\theequation\hbox to
0.6em{\hfil\alph{alphaequation}\hfil}}
\def\eqnsystem#1{
\def\@eqnnum{{\rm (\thealphaequation)}}
\def\@@eqncr{\let\@tempa\relax \ifcase\@eqcnt \def\@tempa{& & &} \or
  \def\@tempa{& &}\or \def\@tempa{&}\fi\@tempa
  \if@eqnsw\@eqnnum\refstepcounter{alphaequation}\fi
\global\@eqnswtrue\global\@eqcnt=0\cr}
\refstepcounter{equation} \let\@currentlabel\theequation \def\@tempb{#1}
\ifx\@tempb\empty\else\label{#1}\fi
\refstepcounter{alphaequation}
\let\@currentlabel\thealphaequation
\global\@eqnswtrue\global\@eqcnt=0 \tabskip\@centering\let\\=\@eqncr
$$\halign to \displaywidth\bgroup \@eqnsel\hskip\@centering
$\displaystyle\tabskip\z@{##}$&\global\@eqcnt\@ne
\hskip2\arraycolsep\hfil${##}$\hfil& \global\@eqcnt\tw@\hskip2\arraycolsep
$\displaystyle\tabskip\z@{##}$\hfil
\tabskip\@centering&\llap{##}\tabskip\z@\cr}

\def\endeqnsystem{\@@eqncr\egroup$$\global\@ignoretrue} \makeatother

\def\diag{\mathop{\rm diag}}
\newcommand{\mb}[1]{\mbox{\normalsize\boldmath $#1$}}

\def\circa#1{\,\raise.3ex\hbox{$#1$\kern-.75em\lower1ex\hbox{$\sim$}}\,}

\begin{document}
\twocolumn[
\parbox{10em}{\raggedleft \em April 1997\\  \bf hep-ph/9704275}\hfill
\parbox{10em}{\bf IFUP--TH 13/97\\ \bf INFN--FE 03--97}\\ \vspace{1cm}
\centerline{\huge\bf\Red The high-$Q^2$ H{\LARGE\bf ERA} anomaly}
\centerline{\huge\bf and supersymmetric unification}

\bigskip\bigskip\Black
\centerline{\large\bf Riccardo Barbieri, Alessandro Strumia} \vspace{0.3cm}
\centerline{\em Dipartimento di Fisica, Universit\`a di Pisa {\rm and}}
\centerline{\em INFN, sezione di Pisa,  I-56126 Pisa, Italia}\vspace{0.3cm}
\centerline{\large and}\vspace{0.3cm}
\centerline {\large\bf Zurab Berezhiani}\vspace{0.3cm}
\centerline{\em INFN, sezione di Ferrara, I-44100 Ferrara, Italy {\rm and}}
\centerline{\em Institute of Physics, Georgian Academy of Sciences, Tbilisi, Georgia}
\bigskip\bigskip\Blue

\centerline{\large\bf Abstract}
\begin{quote}\large\indent
We discuss an initial condition on the superpotential
couplings of an SU(5) theory
which allows the $R$-violating term $Ld^cQ$
but avoids the simultaneous presence of $LLe^c + d^cd^cu^c$.
This same condition keeps under control the products
of pairs of different
couplings $\lambda_{ijk} L_i d^c_j Q_k$, which are more strongly
constrained by flavour-changing-neutral-current limits.
In our view, this observation makes relatively
more plausible the interpretation of
the high-$Q^2$ {\sc Hera} anomaly, if real, as caused by squark production.
\end{quote}\Black
\vspace{1cm}]

\noindent
\paragraph{1}
It is far too early to say if the anomalous events observed by H1 and
{\sc Zeus} at {\sc Hera}~\cite{Hera} in deep-inelastic $e^+p$ scattering
are due to a statistical fluctuation or to physics beyond the Standard
Model. In any event, their finding has stimulated an intensive
phenomenological discussion~\cite{R,RC,Babu,L,Contact}. Here we are
concerned with the possible interpretation of the anomaly as caused by
the production of a squark~\cite{R,RC,Babu} with violation~\cite{HS}
of $R$-parity~\cite{MSSMR}.

At first sight there are two problems which seem to speak, at least in
our view, against this interpretation, although none of them prevents a
purely phenomenological description of the data: the Flavour Changing
Neutral Current problem and the consistency with supersymmetric
unification. In terms of supersymmetric chiral multiplets,
the relevant coupling is
\begin{equation}\label{eq:1}
\lambda_{ijk} L_i d^c_j Q_k
\end{equation}
where $L_i$, $d^c_j$ and $Q_k$ are superfields of lepton doublets, quark
singlets and quark doublets respectively and $i,j,k$ are generation
indices. Table~1 lists the strongest limits on products of
{\em pairs of different couplings\/} of the type\eq{1}:
this is the FCNC problem caused by tree level sparticle exchanges.
We remind that the {\sc Hera} data, interpreted as caused by
$e^+d\to\tilde{u}_{Lk}$, require a {\em single coupling\/}
$\lambda_{11k}\circa{>} 0.03$~\cite{R,RC,Babu}.

Furthermore, as soon as one allows for $R$-parity
or matter-parity breaking, other couplings than\eq{1} are possible ($u^c$ and
$e^c$ are quark and lepton singlets respectively)
\begin{equation}\label{eq:2}
\lambda'_{ijk} L_i L_j e^c_k + \lambda''_{ijk}d^c_i d^c_j u_k^c
\end{equation}
with $\lambda'_{ijk}$ breaking lepton number, as $\lambda_{ijk}$ in\eq{1},
and $\lambda''_{ijk}$  breaking baryon number. In short, the simultaneous
presence of both (1) and (2) in a generic unified theory with broken
matter parity,
 or at least in those ones that may have a chance to account for the {\sc
Hera} anomaly, is the other source of concern.

\paragraph{2}
We cannot offer a neat solution for these problems. We point out,
however, that there is an initial condition, or a choice of unified
couplings at some large scale, higher than the grand unified scale
$\MGUT$, which gets rid of the unwanted couplings (2) and, at the same
time, satisfies the strongest limits in table 1. Somehow, the flavour
alignment suggested by the FCNC constraints might allow to unify the large
$R$-violating interaction, seemingly required by the interpretation of
the {\sc Hera} anomaly, in a way compatible with proton decay.

We consider an SU(5) theory and we denote, as usual, by $F_i$, $T_i$,
$H$, $\bar{H}$ the three generations of five-plets and ten-plets, and the
$5$ and $\bar{5}$ of Higgs-superfields respectively.  We write the
relevant Yukawa superpotential as
\begin{equation}\label{eq:3}
W=\lambda^F_{ij}(\Sigma)\bar{F}_i T_j \bar{H} +
\lambda^T_{ij}(\Sigma)T_i T_j H +
\lambda_{ijk}(\Sigma)\bar{F}_i\bar{F}_j T_k  
\end{equation}
where $\lambda^F_{ij}$, $\lambda^T_{ij}$ and $\lambda_{ijk}$ are
functions of one or more superfields $\Sigma$, whose vacuum expectation
values break SU(5) down to the SM gauge group at $\MGUT$.  As well known,
a breaking of SU(5) is required in
 $\lambda^F_{ij}(\md{\Sigma})$ to account for the different d-quark
 and charged lepton masses. As we shall see, such a breaking is also
needed in $\lambda_{ijk}$. In\eq{3}, all SU(5)-contractions are left
understood.
$\bar{H}$ is defined as the field whose triplet component gets a heavy
mass together with the triplet in $H$ when SU(5) is broken, whereas the
corresponding SU(2)-doublets remain light or, at least, have a
significant component in the light Higgs doublets
$h_2$, $h_1$.

The key assumption that we make is that the couplings
$\lambda_{ij}^F(\md{\Sigma})$ and $\lambda_{ijk}(\md{\Sigma})$, for any $k$,
are simultaneously diagonal in $i,j$\footnote{To
ensure the vanishing of\eq{2} a weaker condition is the symmetry
under $i\leftrightarrow j$ of the
$\lambda_{ijk}(\md{\Sigma})$. The stronger condition
here is imposed to help, at the same time, the FCNC problem.}.
As an illustrative example, consider an expansion of $W$
in inverse powers of
a mass scale $M$ higher than $\MGUT$,
\begin{eqnarray}\label{eq:4}\nonumber
W &=&\lambda_i^{0F} \bar{F}_i T_i \bar{H} +
\frac{\lambda_i^{1F}}{M}\Sigma\bar{F}_iT_i\bar{H}+\\
&&+\lambda_{ij}^{0T} T_i T_j H +
\frac{\lambda_{ij}^{1T}}{M} \Sigma T_i  T_j H+\\
&&
+\frac{\lambda_{ik}}{M}\Sigma \bar{F}_i \bar{F}_i T_k+\cdots\nonumber
\end{eqnarray}
where $\lambda_i^{0F}$, $\lambda_i^{1F}$, $\lambda_{ij}^{0T}$,
$\lambda_{ij}^{1T}$ and
$\lambda_{ik}$ are all dimensionless couplings and $\Sigma$ is an
SU(5)-adjoint getting a non-zero vacuum expectation value. Notice that
there is no cubic term $\bar{F}_i\bar{F}_i T$ in\eq{4}, since the
$\bar{5}\otimes\bar{5}\otimes10$ SU(5)-invariant coupling is antisymmetric
under interchange of the two $\bar{5}$'s.

Our simple observation is that the last term on the right-handed side
of\eq{3}, or\eq{4}, below the unification scale,
after $\Sigma\to\md{\Sigma}$, reduces to a term proportional to
\begin{equation}\label{eq:5}
\lambda_{ik} L_i d_i^c Q_k
\end{equation}
whereas the terms of the form\eq{2} automatically vanish due to the
antisymmetry of such couplings in the indices $i,j$. At the
same time, the simultaneous diagonality of the couplings
$\lambda_{ij}^F(\md{\Sigma})$ and
$\lambda_{ijk}(\md{\Sigma})$ reduces to zero all the
pair of couplings in table~1 except for the one relevant to
$K_L\to\mu e$, which remains a product of two free parameters. We are
neglecting here possible modifications of the form of the
superpotential\eq{4} due to loop corrections (see below). As we said, we
cannot really claim a solution for the difficulties pointed out in the
introduction.
Still, the fact that one can get around them by a simple initial
condition on the unified couplings may not be accidental.


\begin{table}
$$\begin{array}{ccrcl}
\varepsilon_K&: & \Im\lambda_{i12}^{\phantom{*}}
\lambda^*_{i21}\cl 3\cdot 10^{-11}\\
\Delta m_K&: & \lambda_{i12}\lambda_{i21}\cl 5\cdot 10^{-9}\\
\mu\hbox{Ti}\to e\hbox{Ti}&:&\lambda_{11k}\lambda_{21k}\cl2\cdot10^{-7}\\
\mu\hbox{Ti}\to e\hbox{Ti}&:&\lambda_{1j1}\lambda_{2j1}\cl2\cdot10^{-7}\\
\Delta m_B&:& \lambda_{i13}\lambda_{i31}\cl3\cdot 10^{-7}\\
K_L\to\mu e&:&\lambda_{11k}\lambda_{22k}\cl3\cdot 10^{-6}
\end{array}$$
\caption{\em Limits on products of two different couplings
$\lambda_{ijk} L_i d_j^c Q_k$. All bounds scale as
$(m_{\tilde{f}}/200\GeV)^2$, with $m_{\tilde{f}}$ the mass of the
relevant exchanged scalar.}
\end{table}

\paragraph{3} If we neglect some small, but interesting, renormalization
effects which modify the form of the superpotential\eq{4}, to be
discussed below, the superpotential\eq{4} reduces at low energy to
\begin{equation}\label{eq:6}
W_{\rm l.e.}=
h_2 u_c^T\mb{\lambda}^{\rm u}Q + h_1d_c^T \mb{\lambda}^{\rm d}Q +
h_1e_c^T \mb{\lambda}^{\rm e} L+L_i d_c^T \mb{\lambda}^{(i)} Q
\end{equation}
where we have introduced a matrix notation in flavour space. In the same
basis as\eq{4}, $\mb{\lambda}^{\rm u}$ is an arbitrary matrix,
$\la{d}$ and $\la{e}$ are both diagonal and\footnote{Trivial rescaling
factors are reabsorbed by proper redefinitions of the parameters.}
\begin{equation}\label{eq:7}
\mb{\lambda}^{(i)}_{jk}=\delta_{ij} \lambda_{ik}
\end{equation}
 For the purposes of this discussion we also assume a flavour universal
initial condition for the supersymmetry breaking sfermion
masses.

By going to the physical basis also for the u-quarks,
the $R$-parity violating interaction\eq{5}
becomes
\begin{equation}\label{eq:8}
e_i d_c^T\mb{\lambda}^{(i)} \mb{V}^\dagger u -
\nu_i d_c^T \mb{\lambda}^{(i)} d
\end{equation}
where $\mb{V}$ is the usual Cabibbo-Kobayashi-Maskawa matrix.

The first term in\eq{8} is responsible for squark production at {\sc
Hera}. Restricting ourselves to valence quark production, all the
$\tilde{\rm u}$-squarks are necessarily produced at {\sc Hera} if
kinematically accessible, via\footnote{Here we neglect
$\tilde{t}_L/\tilde{t}_R$ mixing.}
$$e^+ d_R\to\tilde{u}_L,\tilde{c}_L,\tilde{t}_L,$$
with effective couplings
$\lambda_{11k}V_{1k}^*$, $\lambda_{11k}V_{2k}^*$ and
$\lambda_{11k}V_{3k}^*$ respectively.
To explain the {\sc Hera} anomaly by any one of these production
processes, for a
$\tilde{\rm u}$-squark of about $200\GeV$ in mass, the effective coupling
must be about $0.04/\sqrt{\hbox{B}}$ \cite{R,RC}, where B{} is the
branching ratio for the same squark into the $R$-violating mode
$e^+ u$. On the other hand, the exchange of
the $\tilde{u}_L$ squark gives rise to
an unobserved neutrinoless $\beta\beta$ decay in Ge
unless~\cite{betabeta}
\begin{equation}
\left| \lambda_{11k} V_{1k}^*
\left(\frac{200\GeV}{m_{\tilde{u}_L}}\right)^{\!\!2}
\right|<7\cdot 10^{-3}
\left(\frac{M_3}{1\TeV}\right)^{\!1/2}
\end{equation}
where $M_3$ is the gluino mass and $m_{\tilde{u}_L}$ the mass of
$\tilde{u}_L$. Let us consider only one of the three couplings
$\lambda_{11k}$ at a time. Taking into account the values of the CKM
matrix elements, to explain the {\sc Hera} anomaly,
$\lambda_{111}$ (dominant $\tilde{u}_L$ production) is excluded,
$\lambda_{112}$ (dominant $\tilde{c}_L$ production) is at the border, in
any case with a branching ratio B{} close to unity and with a
$\beta\beta$ decay signal around the corner, whereas
$\lambda_{113}$ (dominant $\tilde{t}_L$ production) is certainly
possible\footnote{Note that a stop significantly lighter than
$\tilde{u}_L$, $\tilde{c}_L$ is motivated in supergravity models}.

\paragraph{4} A crucial question that we have to address is to what
extent the form of the superpotential\eq{4}, which must be viewed as an
initial condition valid at some large scale $M$, maybe close to the
Planck scale or the string scale, is stable under renormalization. Since
it is not, this question is actually of interest, {\em mutatis
mutandis\/}, for any attempt to explain the {\sc Hera} anomaly by
$R$-parity breaking.

\smallskip

The main tool here is the non-renormalization theorem~\cite{NRT} which
states that the superpotential\eq{3}, or\eq{4}, undergoes only
wave-function renormalization. Notice that wave-function renormalization
does produce mixing, in general, which means that the flavour structure
of the superpotential\eq{4} will not be maintained. On the other hand,
wave-function renormalization cannot generate a term which is not there
to start with {\em for any flavour structure\/}. This shows that terms of
the form\eq{2}, possibly with an extra factor containing some component
of the $\Sigma$-field, will never arise. The argument applies both to
renormalization effects above and below the unification scale.

To calculate the modification of the coupling\eq{8} both from
renormalization above and below the unification scale, it is convenient
to go to the basis where the Yukawa coupling matrix of the u-quarks,
$\lambda^T_{ij}$ in eq.\eq{4}, is diagonal. This is because we
concentrate on the effects of the large top quark Yukawa coupling
$\lambda_t$~\cite{RGE}. In this basis, above the unification scale, the
three ten-plets $T$ are rescaled by
\begin{equation}
T\to\diag(1,1,y_{\rm G}) T\equiv \mb{Y}_{\rm G} T
\end{equation}
where
$$y_{\rm G} \equiv \exp \bigg[-3\int_{\MGUT}^M
\lambda_t^2(E)\frac{d\ln E}{(4\pi)^2} \bigg]$$ so that the u-quark
Yukawa coupling matrix stays diagonal and will ever remain so even below
$\MGUT$. On the other hand, at $\MGUT$, the full superpotential acquires
the form
\begin{eqnarray}\nonumber W_{\rm G}&=& HT^T\la{T}_{\rm G} T +
\bar{H}\bar{F}^T\la{F}_{\rm G}(\Sigma)\mb{V}_{\rm G}^\dagger \mb{Y}_{\rm
G} T+\\ &&+
\bar{F}_i~\bar{F}\mb{\lambda}^{(i)}_{\rm G}(\Sigma)
\mb{V}_{\rm G}^\dagger
\mb{Y}_{\rm G} T
\end{eqnarray} where $\mb{V}_{\rm G}$ is the unitary rotation that
diagonalizes $\la{u}$.

Just below the unification scale, $W_{\rm G}$ must be restricted to the
massless fields, becoming
\begin{eqnarray*} W_{{\rm below}~\MGUT}&=& h_2 u_c^T \la{u}_{\rm G} Q +
h_1 d_c^T \la{d}_{\rm G} (\mb{V}_{\rm G}^\dagger \mb{Y}_{\rm G}) Q+\\
&&+h_1 L^T (\la{e}_{\rm G} \mb{V}_{\rm G}^\dagger \mb{Y}_{\rm G}) e_c +\\
&&+ L_i~d_c^T(\mb{\lambda}^{(i)}_{\rm G} \mb{V}_{\rm G}^\dagger
\mb{Y}_{\rm G}) Q
\end{eqnarray*} From $\MGUT$ to the Fermi scale, the interesting effect
is the further rescaling of the $Q$ fields
\begin{equation} Q\to\diag(1,1,y_t) Q\equiv \mb{Y}_t Q
\end{equation}
with
$$y_t \equiv \exp\bigg[ -\int^{\MGUT}_{M_Z}
\lambda_t^2(E)\frac{d\ln E}{(4\pi)^2}\bigg], $$
thus
reducing the superpotential at low energy
to a modified form with respect to\eq{6}
\begin{eqnarray}\nonumber
W^{\rm true}_{\rm l.e.} &=& h_2 u_c^T \la{u}  Q +\\ 
&&+\nonumber
h_1 d_c^T(\la{d}_{\rm G} \mb{V}_{\rm G}^\dagger
\mb{Y}_t \mb{Y}_{\rm G}) Q+\\ \label{eq:14}
&&+h_1 L^T(\la{e}_{\rm G} \mb{V}_{\rm G}^\dagger
\mb{Y}_{\rm G}) e_c + \\  \nonumber
&&+L_i~d_c^T (\mb{\lambda}^{(i)}_{\rm G}
\mb{V}_{\rm G}^\dagger \mb{Y}_t\mb{Y}_{\rm G}) Q
\end{eqnarray}
The rescaling due to the gauge couplings, being flavour independent, does
not concerns us here and can be reabsorbed in an overall redefinition of
the various couplings.

\medskip

To have a physical interpretation of\eq{14},
one has to go to the physical basis both for the d-quarks
and for the charged leptons (the u-quark mass matrix is already diagonal).
This is achieved by proper unitary rotations defined by
\begin{eqnsystem}{sys:15}
\la{e}_{\rm G} \mb{V}_{\rm G}^\dagger \mb{Y}_{\rm G}  &\equiv&
\mb{U}_L^\dagger \la{e} \mb{V}_L^\dagger\\
\la{d}_{\rm G} \mb{V}_{\rm G}^\dagger \mb{Y}_t\mb{Y}_{\rm G}  &\equiv&
\mb{U}^\dagger \la{d} \mb{V}^\dagger\label{eq:15b}
\end{eqnsystem}
where $\la{e}$ and $\la{d}$ are the diagonal low-energy Yukawa couplings
of the charged leptons and d-quarks respectively.
In\eq{15b}, $\mb{V}$ is the CKM matrix, which justifies
the use of the same notation as in\eq{8}.
By going to the physical basis, the true form of the $R$-violating
couplings, rather than\eq{8}, becomes therefore
\begin{equation}\label{eq:16}
e_i U_{Lij} d_c^T \mb{\lambda}^{(i)}_{\rm true} \mb{V}^\dagger u-
\nu_i U_{Lij} d_c^T \mb{\lambda}^{(i)}_{\rm true}d
\end{equation}
where
\begin{equation}\label{eq:17}
\mb{\lambda}^{(i)}_{\rm true} = \mb{U}
\mb{\lambda}^{(i)}_{\rm G}\frac{1}{\la{d}_{\rm G}}
\mb{U}^\dagger \la{d}
\end{equation}
Notice that, except for the original $\mb{\lambda}^{(i)}_{\rm G}$
matrices,
all other matrices in\eq{16},\eq{17} are known.
By solving eq.s~(\ref{sys:15}) at leading order in ratios of small Yukawa
couplings one finds
($\la{d}=\diag(\lambda_d,\lambda_s,\lambda_b)$)
\begin{eqnarray}\nonumber
\la{d}_{\rm G} &=& \diag(\lambda_d,\lambda_s,\lambda_b/y_t)\\
\label{sys:rge}
|U_{13}| &=&|U_{31}| =(1-y^2)(\lambda_d/\lambda_b) |V_{td}| \\
\nonumber
|U_{23}|&=&|U_{32}| =(1-y^2)(\lambda_s/\lambda_b) |V_{ts}|\\
\nonumber
|U_{12}| &=&|U_{21}| = (1-y^2)(\lambda_d/\lambda_s) |V_{td} V_{ts}^*|
\end{eqnarray}
where $y = y_t y_{\rm G}$. $\mb{U}_L$ has the same form as $\mb{U}$,
with $\la{d}$ replaced by $\la{e}$ and $y$ by $y_{\rm G}$.

\smallskip

The upshot of all this is that, even if one starts at high energy with
couplings
$\mb{\lambda}^{(i)}$ of the form\eq{7}, small rotations occur both on the
$L_i$-index (only due to GUT effects) and on the $d_i^c$. In particular,
it is no longer true that the pairs of couplings occurring in table~1
(all but the one for $K_L\to \mu e$) identically vanish. Even individual
lepton numbers are broken by renormalization effects above the
unification scale. However, taking into account of
eq.~(\ref{eq:16}$\div$\ref{sys:rge}), these flavour effects are small
enough. Whatever choice is made of the index $k=2,3$ in $\lambda_{11k}$
to explain the {\sc Hera} anomaly, none of the bounds in table~1 is
violated. The model passes this consistency check.

\paragraph{5} Nothing has been said so far on the supersymmetry breaking
terms, except that the sfermion masses are taken diagonal in the basis
defined by\eq{4}. If $A$-terms are generated by supergravity
couplings~\cite{SuGraSoft}, we assume that their flavour structure at the
Planck scale is the same as in\eq{4}. As such, their discussion is
analogous to the one for the superpotential itself. In particular, no
baryon number violating $A$-term,
$\tilde{d}^c_i\tilde{d}^c_j\tilde{u}^c_k$, is there to start with, nor it
is generated by radiative corrections, up to terms that vanish at least
as $m/\MGUT$, where $m$ is the low energy effective supersymmetry
breaking scale. Such terms will ultimately give rise to proton decay at a
rate proportional to $\MGUT^{-2}$, as from a dimension-5 baryon-number
violating operator~\cite{dim5p-decay}, but with a highly suppressed
numerical coefficient.

\medskip

Neutrino masses also require a discussion. We assume that the $\mu$-term
and the $B$-term, generated after supersymmetry breaking by supergravity
couplings, do not involve, as an initial condition, the fields
$\bar{F}_i$, but only the light fragments of $H$ and $\bar{H}$. The
theory knows the difference between $\bar{F}_i$ and $\bar{H}$, since, by
definition, it is the triplet in $\bar{H}$ which becomes heavy after
SU(5)-breaking. In this situation no neutrino mass is present at the tree
level. All the three neutrinos will receive mass, however, after
radiative corrections. We have checked that none of them, for natural
values of the parameters, exceeds the level of $1\eV$.

\paragraph{6} In conclusion we have discussed a
simple idea on the initial conditions for the 
superpotential couplings of an SU(5) theory which allows the 
$R$-violating terms\eq{1} but avoids the simultaneous presence of 
the terms\eq{2}.
This same condition makes the products of pairs of different couplings of
the form\eq{1} to vanish, which are more
strongly constrained by FCNC limits. In
our view, this makes relatively
more plausible the interpretation of the high-$Q^2$
{\sc Hera} anomaly, if real, as caused by squark production. We have
pointed out some phenomenological consequences of our hypothesis, some of
which require further study. Although possible, we do not find useful to
speculate, at this time, on a symmetry origin for such initial condition
at the Planck scale.

\paragraph{Acknowledgements}
We thank Francesco Vissani and Misha Vysotsky
for useful discussions.

\end{document}